\documentclass[10pt,twocolumn,letterpaper]{article}

\usepackage{iccv}
\usepackage{times}
\usepackage{epsfig}
\usepackage{graphicx}
\usepackage{amsmath}
\usepackage{amssymb}
\usepackage{float}
\usepackage{comment}
\usepackage{subcaption}
\usepackage{booktabs}
\usepackage{threeparttable,booktabs}
\usepackage{longtable}
\usepackage[hang,flushmargin]{footmisc}
\usepackage{multirow}

\usepackage[breaklinks=true,bookmarks=false]{hyperref}

\iccvfinalcopy 

\def\iccvPaperID{1804} 

\ificcvfinal\fi

\begin{document}

\title{MedKLIP: Medical Knowledge Enhanced Language-Image \\ Pre-Training in Radiology}

\author{Chaoyi Wu\textsuperscript{1,2}, \quad Xiaoman Zhang\textsuperscript{1,2}, \quad Ya Zhang\textsuperscript{1,2},
\quad Yanfeng Wang\textsuperscript{1,2}, \quad Weidi Xie\textsuperscript{1,2} \\
\textsuperscript{1}Cooperative Medianet Innovation Center, Shanghai Jiao Tong University \quad
\textsuperscript{2}Shanghai AI Laboratory \quad\\
\tt\small{\{wtzxxxwcy02, xm99sjtu, ya\_zhang, wangyanfeng, weidi\}@sjtu.edu.cn}\\
\url{https://chaoyi-wu.github.io/MedKLIP/}
}

\maketitle
\ificcvfinal\thispagestyle{empty}\fi

\begin{abstract}
   In this paper, we consider enhancing medical visual-language pre-training~(VLP) with domain-specific knowledge, 
   by exploiting the paired image-text reports from the radiological daily practice.
   In particular, we make the following contributions:
    {\bf First}, unlike existing works that directly process the raw reports,
    we adopt a novel triplet extraction module to extract the medical-related information, avoiding unnecessary complexity from language grammar and enhancing the supervision signals;
   {\bf Second}, we propose a novel triplet encoding module with entity translation by querying a knowledge base, to exploit the rich domain knowledge in medical field, and implicitly build relationships between medical entities in the language embedding space;
   {\bf Third}, we propose to use a Transformer-based fusion model for spatially aligning the entity description with visual signals at the image patch level, enabling the ability for medical diagnosis;
   {\bf Fourth}, we conduct thorough experiments to validate the effectiveness of our architecture, 
   and benchmark on numerous public benchmarks {\em e.g.}, ChestX-ray14, 
   RSNA Pneumonia, SIIM-ACR Pneumothorax, COVIDx CXR-2, COVID Rural, and EdemaSeverity. In both zero-shot and fine-tuning settings, our model has demonstrated strong performance compared with the former methods on disease classification and grounding.
\end{abstract}

\section{Introduction}
\label{sec:intro}

\begin{figure}
    \centering
    \includegraphics[width=\linewidth]{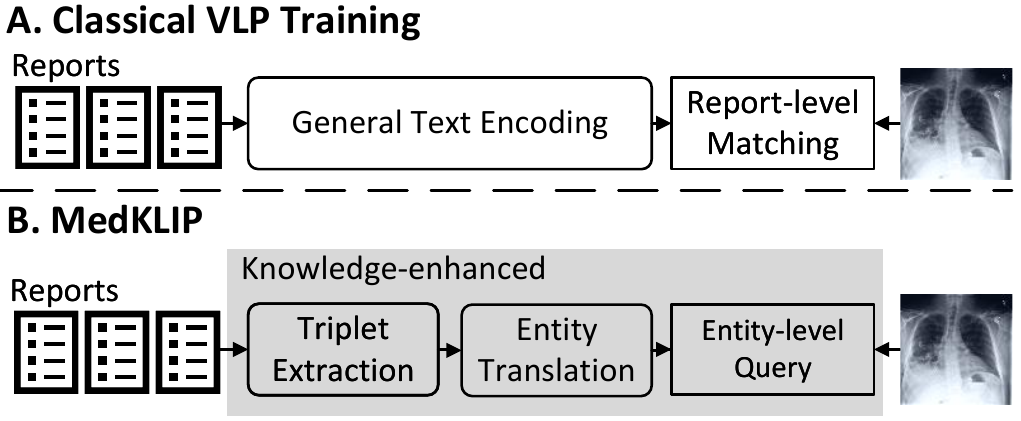}
    \caption{Our method mainly considers combining medical knowledge with VLP. We propose Triplet Extraction and Entity Translation modules, 
    so that the network can be supervised with detailed entity-level signals.}
    \label{fig:intuition}
    \vspace{-0.5cm}
\end{figure}

With the rapid development of deep learning, 
numerous works have been proposed to facilitate computer-aided diagnosis in the medical field~\cite{miura2020improving,esteva2017dermatologist,titano2018automated,erickson2017machine}. Despite the tremendous progress, these models are normally trained to recognize or segment the structures that fall into a certain closed set of anatomical or disease categories, 
whenever a new disease comes to be of interest, a costly procedure for data annotation, model re-training are required, 
fundamentally limiting its practical values.
As an alternative, recent research considers to train the model on the corpus, consisting of large amount of multi-modal data, that is generated from daily clinical routine, for instance, the most common example is the dataset of X-ray images with paired radiological reports~\cite{dunnmon2020cross,irvin2019chexpert,johnson2019mimic}. 


This paper presents our preliminary investigation on vision-language representation learning in the medical domain, with the goal of better zero-shot disease diagnosis (classification) and grounding. Undoubtedly, these tasks have also been widely investigated in the computer vision community, with significant progress made on developing Foundational Models in the past years, for example, CLIP~\cite{radford2021learning}, ALBEF~\cite{li2021align}, BLIP\cite{li2022blip}, etc. However, to achieve such a goal in the medical domain, 
different challenges must be resolved, 
that requires research efforts from the community: {\em First}, data availability, 
training Foundation Models in computer vision normally require over millions of image-text pairs, while in the medical domain, 
only a few hundred thousand pairs are available~\cite{johnson2019mimic}. 
The limited data challenges language models to understand the reports in free form~\cite{boecking2022making}.
{\em Second}, the problem considered in computer-aided diagnosis is naturally fine-grained,
that requires distinguishing the medical concepts to understand the disease, 
as a consequence, domain knowledge is essential;
{\em Third}, robustness is crucial, it is, therefore, preferable to have explainability, 
where diagnosis results come along with the visual grounding, to help radiologists understand the system, and build trust between human and machines.

Existing work in medical VLP~(Vision-Language Pre-training)~\cite{zhang2020contrastive,muller2021joint,huang2021gloria,boecking2022making} follows a straightforward training paradigm by matching raw reports with image scans, 
as shown in Fig.\ref{fig:intuition}A, ignoring the medical prior knowledge, and, thus, we propose a novel knowledge-enhanced visual-language model as shown in Fig.~\ref{fig:intuition}B.
{\em First}, we propose a triplet extraction module to extract useful medical entities~(keywords) from raw reports, and simplify each report into sets of triplets, 
denoted as $\{{\rm entity, position, exist}\}$. Decomposing reports into triplets leads to an effective representation of the reports with minimal information loss due to the structural prior in reports; 
{\em Second}, we translate the medical entities into fine-grained descriptions by leveraging a well-defined medical word knowledge base, that tends to explain diseases with common vocabulary. 
Thus, computing text embeddings for these descriptions enables to implicitly establish relationships between medical entities;
{\em Third}, we view the entities as a query set and adopt a transformer-based architecture for aligning the image patches with entity descriptions, that enables explicit supervision signals at entity level. Consequently, we can simultaneously infer the likelihood of certain diseases with the visual evidence in the form of a spatial heatmap, {\em i.e.}, providing rough grounding for explainability. 

We pre-train the model on one widely-used medical image-report dataset MIMIC-CXR~\cite{johnson2019mimic},
and rigorously evaluate on the task of disease diagnosis across numerous public benchmarks, 
{\em e.g.}, ChestX-ray14~\cite{wang2017chestx}, RSNA Pneumonia~\cite{shih2019augmenting}, SIIM-ACR Pneumothorax~\cite{kaggle-siim}, COVIDx CXR-2~\cite{pavlova2021covid}, COVID Rural~\cite{tang2020deep,desai2020chest}, and EdemaSeverity~\cite{chauhan2020joint}. 
We get state-of-the-art performance on zero-shot classification and grounding on different diseases, spanning different image distributions, 
with further fine-tuning, our model still exceeds previous models significantly.

\section{Related Work}
\noindent \textbf{General Vision-Language Pre-training~(VLP) Models.}
    In computer vision, such line of research has gained tremendous success in the recent literature, generally speaking, the developed architectures can either be two-stream~\cite{bianchi2021contrastive,li2021align,jia2021scaling}, 
    {\em i.e.}, dual encoders, or those based on single-stream methods~\cite{li2019visualbert,chen2020uniter}, that favors visual-language fusion. 
    In particular, several works~\cite{cui2021rosita,li2020oscar,yu2021ernie} consider considers to combine the commonsense knowledge into the vision-language pre-training, however, in this paper, we focus on medical domain, which is clearly more fine-grained and requires significantly more expertise.
    

    

\vspace{5pt}
\noindent \textbf{Medical Named-Entity-Recognition~(NER) Models.}
    Various natural language processing~(NLP) approaches have been proposed to extract information from radiology reports~\cite{peng2018negbio,irvin2019chexpert,mcdermott2020chexpert++,smit2020chexbert}. These early methods considered only the disease, 
    thus causing information loss.
    Further state-of-the-art works~\cite{jain2021radgraph,wu2021chest} are proposed to extract relationship between different entities without demand of pre-defined close disease set, retaining most of useful information with high accuracy. In weakly supervision~\cite{yu2022anatomy} and report generation fields~\cite{delbrouck2022improving}, NER methods have shown great impact, 
    and greatly inspired us for more effective vision-language pre-training with medical domain knowledge injected.

\vspace{2pt}
\noindent \textbf{Medical Knowledge Enhanced Models.}
    Leveraging external medical knowledge to enhance deep learning models is not a new topic~\cite{xie2021survey}. Depending the approaches of using medical knowledge, They can be classified into model-based or input-based. 
    In model-based approaches, 
    the authors aim to imitate the radiological or diagnosis practice to design models~\cite{li2019attention,gonzalez2018dermaknet,wang2020learning,huang2020dual,mitsuhara2019embedding,fang2019attention,cui2020collaborative}. 
    While in input-based approaches, the knowledge is treated as an extra input for computing features~\cite{yang2019dscgans,xie2018knowledge,tan2019expert,chen2020generating} or to guide the final training loss~\cite{chen2016automatic, hussein2017risk,li2019canet,liao2019multi,maicas2018training}, 
    commonly used in report generation tasks~\cite{wang2022medical,yang2022knowledge,liu2021exploring,chen2020generating}. However, none of these works are targeting on vision-language pre-training in medical domain with image-report.
    
    
\vspace{2pt}
\noindent \textbf{Concurrent Works in Medical VLP.}
Existing medical VLP methods follow the two-stream flow~\cite{zhang2020contrastive,huang2021gloria,chauhan2020joint,muller2021joint,wang2022medclip,chen2022align}, {\em i.e.}, use contrastive learning and without fusion module, for example,
ConVIRT~\cite{zhang2020contrastive} initially proposed to use contrastive loss as a proxy task for aligning the medical scan and corresponding reports,
LoVT and GLoRIA then focus on improving the local alignment performance~\cite{huang2021gloria,chauhan2020joint}. 
BioViL notices the language pattern in reports is different from natural texts and re-designs the language model~\cite{boecking2022making}. 
Align~\cite{chen2022align} improves the performance of VLP model on medical Visual Question Answering~(VQA). 
The recent arxiv preprint, MedCLIP~\cite{wang2022medclip}, 
considers leveraging unpaired data to make up data scarcity. The most related to ours is CheXzero~\cite{tiu2022expert}, which targets at zero-shot diagnosis. Despite significant contribution has been made by existing work~\cite{zhang2020contrastive,huang2021gloria,chauhan2020joint,muller2021joint,wang2022medclip,tiu2022expert}, 
they still treat medical texts and images as common natural data and do not explicitly leverage the rich prior knowledge from medical domain. 
In this paper, we consider to incorporate domain knowledge, 
re-design the pre-training pipeline delicately and target at accurate diagnosis in X-ray scans.


\begin{figure*}[t]
    \centering
    \includegraphics[width=\textwidth]{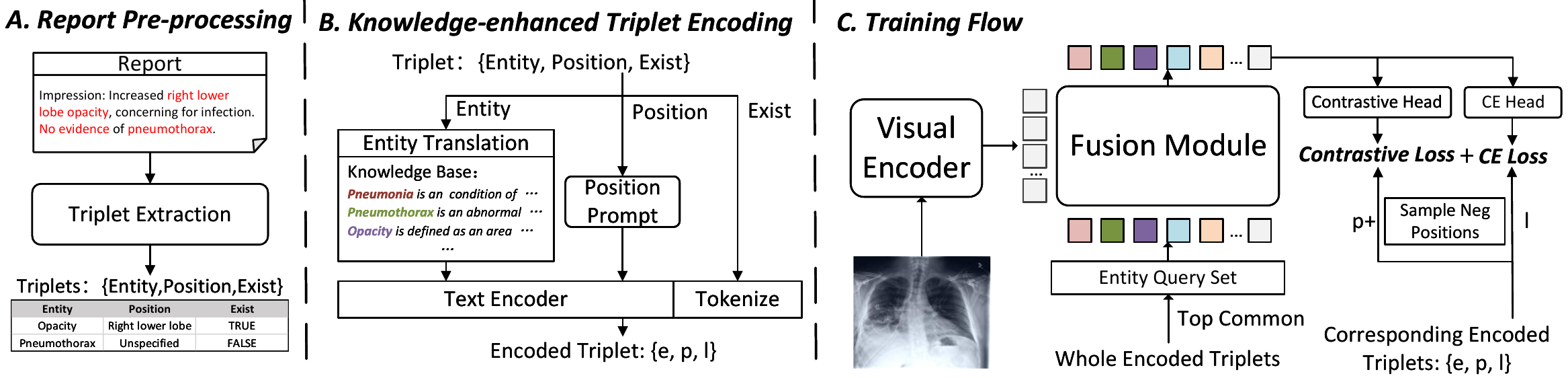}
    \caption{The whole framework of our method. We first pre-process the report into triplets leveraging triplet extraction module. Then we encode the extracted triplets and it is worth emphasizing that we translate the entities into detailed descriptions during encoding, by querying the medical knowledge base. Finally we change the training flow with triplets, \emph{i.e.}, we query a transformer-based fusion module at entity level, which provides more detailed supervision signals.
    }
    \label{fig:network}
    \vspace{-0.5cm}
\end{figure*}

\section{Method}
In this section, we start by describing the considered problem scenario in Sec.~\ref{sec:problem_scenario}, 
followed by our report pre-process operation with triplet extraction  in Sec.~\ref{sec: Pre-processing}. Then we introduce our proposed knowledge-enhanced architecture in Sec.~\ref{sec:architecture}, including, 
visual encoder, knowledge-enhanced triplet encoder, 
and the fusion module for aligning visual-language signals.
In Sec.~\ref{sec:training}, we describe the training procedure with the paired image-reports sourced from the daily routine X-ray scans and, in Sec.\ref{sec: infer}, we introduce the procedure for inference.

\subsection{Problem Scenario}
\label{sec:problem_scenario}
Assuming we are given a training set with $N$ samples, 
{\em i.e.}, $\mathcal{D}_{\text{train}} = \{(\mathcal{X}_1, \mathcal{T}_1), \dots, (\mathcal{X}_N, \mathcal{T}_N)\}$,
where $\mathcal{X}_i, \mathcal{T}_i$ refer to the X-ray image and its corresponding medical report generated in the daily routine scans, respectively,
our goal is to train a visual-language model that enables us to diagnose the existence of certain diseases and localize the visual evidence spatially.
Specifically, at inference time, we can freely ask the system to identify the likelihood of the patient getting a certain disease~(seen or unseen during training): 
\begin{align}
    \hat{s}_i, \hat{m}_i = \Phi_{\text{fusion}}(\Phi_{\text{visual}}(\mathcal{X}_i), \Phi_{\text{textual}}(\text{[description]})),
\end{align}
where $\mathcal{X}_i \in \mathbb{R}^{H \times W \times 3}$ refers to an image sample from the \textbf{test set},  with $H, W$ denoting height and width respectively. 
$\hat{s}_i \in [0, 1]$ refers to the inferred likelihood of the patient having a certain disease indicated by the input description, 
and $\hat{m}_i \in \mathbb{R}^{H \times W \times 1}$ denotes a predicted spatial heatmap, with high activation on pixels that potentially provide the visual indication for such disease. 
In the following section, we will introduce our report pre-process operation with triplet extraction.


\subsection{Report Pre-processing}
\label{sec: Pre-processing}
To start with, we propose to pre-process medical reports with a \textbf{Triplet Extraction} module by removing the unnecessary complexity from language grammar. Note that, we hereon only consider single sampled image-reports pair $(\mathcal{X}_i, \mathcal{T}_i)$, and ignore the subscript in notations for simplicity.

We condense the original reports with an off-shelf medical Named Entity Recognition~(NER) method, namely RadGraph~\cite{jain2021radgraph,yu2022anatomy}, 
transforming reports into a set of triplets, as shown in Figure~\ref{fig:network}A.
In detail, the medical key words can be extracted and classified as ``entity'' or ``position'' with the NER module.  ``Entity'' refers to some clinical observations, like ``Opacity''. ``Position'' refers to the anatomical body part that occurs in a radiology report, 
like “right lower lobe”. 
Besides, the NER module will also provide an ``exist'' label to conclude whether an entity is claimed to be exist, absent or uncertain in reports. Based on this, we can use a set of triplets, \emph{i.e.}, $\{\text{entity}, \text{position}, \text{exist}\}$, 
to re-formulate the sentence in reports, 
for example,
the triplet $\{\text{Opacity}, \text{Right lower lobe}, \text{True}\}$ represents ``It is true that there is opacity located at right lower lobe''. 
{\bf Note that}, the triplets with a specific ``position'' are not always termed as True in ``exist'' as radiologists may point out entities absent at some specific position.


Therefore, given a report ${ \mathcal{T}}$ with multiple sentences, 
${ \mathcal{T}}=\left \{s_1,s_2,...,s_M\right\}$, 
the extraction module independently operates on each of the sentences,
and construct a number of triplets from the report:
\begin{align}
   \Phi_{\text{ex}}({s_j}) = \{\text{entity}_n, \text{position}_n, \text{exist}_n\}, n \in [0,t_j],
\end{align}
where $t_j$ represents the total number of entities contained in one sentence, 
with $n=0$ indicating the special case that there is no valid entity. 
After the triplet extraction, each report is equal to a set of triplets.

\vspace{3pt}
\noindent  \textbf{Discussion.} 
In contrast to natural texts, 
information in medical reports tends to be more condensed, 
with radiologists pointing out the existence of abnormality and their positions in the image. 
Meanwhile, medical terminologies tend to be professional, 
and within certain vocabulary~(mostly listed in UMLS~\cite{bodenreider2004unified}), 
specially designed NER methods~\cite{jain2021radgraph} demonstrate great performance on reports.
Therefore, adopting the {\bf Triplet Extraction} operation in medical VLP can avoid unnecessary complexity from understanding grammar, 
while still retaining the useful information in reports.

\subsection{Architecture}
\label{sec:architecture}

In this section, 
we detail our proposed framework, consisting of three components,
namely, visual encoding, knowledge-enhanced triplet encoding, 
and fusion module, as shown in Fig.~\ref{fig:network}B and Fig.~\ref{fig:network}C. 


\subsubsection{Visual Encoding}
Given an X-ray image scan $\mathcal{X} \in \mathbb{R}^{H \times W \times 3}$, 
we can compute the features with a visual backbone:
\begin{align}
    \mathcal{V} = \Phi_{\text{visual}}(\mathcal{X}) \in \mathbb{R}^{h \times w \times d},
\end{align}
$h,w,d$ refer to the height, width, 
and feature dimension of the output,
in our case, we adopt ResNet-50 as the visual backbone,
and take the output from the 4th residual block.
Note that, we make the architectural choice for fair comparison with existing work~\cite{zhang2020contrastive,muller2021joint,huang2021gloria,boecking2022making}, 
while other visual backbones, {\em e.g.}, ViT~\cite{dosovitskiy2020image}, 
can equally be applied.

\subsubsection{Knowledge-enhanced Triplet Encoding}
The goal of this module is to encode the triplets extracted from reports by incorporating medical domain knowledge as shown in Fig.\ref{fig:network}B.

Given a triplet as $\{ \text{entity}, \text{position}, \text{exist} \}$, 
it is easy to code the ``exist'' as it only has three outcomes. 
We use $l\in\{0,1,-1\}$ to tokenize it, $1$ for True, $0$ for False, $-1$ for uncertain. 
For the ``entity'' words, 
we translate them into \textbf{detailed descriptions} by querying some easy-access medical knowledge bases~\footnote{Wikipedia~\url{https://en.wikipedia.org/wiki/}}\footnote{UMLS~\cite{bodenreider2004unified}~\url{https://www.nlm.nih.gov/research/umls/}}, \emph{e.g.}, 
Description(\text{[``Pneumonia'']})=``It is a condition of the lung primarily \dots present with opacities and pleural effusion \dots''.
Despite its simplicity, converting the entities into descriptions is crucial for more reliable and zero-shot diagnosis, as it further decomposes the professional medical entities into basic attributes that are shared by different diseases, encouraging the model to capture a deep understanding of the visual evidence. For the ``position'' words, we use a prompt as ``It is located at \{position\}'' to form a sentence. 
Finally, we use ClinicalBERT~\cite{alsentzer2019publicly} as a pre-trained text encoder,
to compute the embedding for the ``entity'' and ``position'', 
and then adopt a linear MLP to project the embedding to desired dimensions:
\begin{align}
    &e = \Phi_{\text{textual}}(\text{Description(\text{\{entity\}) })} \in \mathbb{R}^d, \\
    &p =  \Phi_{\text{textual}}(\text{``It is located at  \{position\}''}) \in \mathbb{R}^{d^{\prime}}.
\end{align}
Each triplet has now been embedded into $\{e, p, l\}$.

\vspace{3pt}
\noindent \textbf{Discussion.}
The extracted entities are medical terminologies 
that are only understandable to audiences with a medical background,
while enriching them with detailed descriptions helps the model to capture a deep understanding of visual evidence for diseases.
Such patterns can be generalized across diseases, as many attribute descriptions tend to be shared, enabling the model to build implicit relationships on seen classes and understand descriptions for unseen ones.


\subsubsection{Fusion Module}
With the triplets from reports, 
we can supervise the model \textbf{on the entity level} instead of the entire report level. The ``position'' and ``exist'' parts in triplets can be naturally seen as more fine-grained supervision labels. Specifically, we adopt a Transformer-based architecture,
use the embedding of entities as query, 
iteratively attending the image embeddings, 
and output exist and position predictions of entities.

In detail, we select the top $|Q|$ most commonly appearing entities' embeddings in all training reports, 
to form an entity query set $Q = \{e_1, e_2,...,e_{|Q|}\}$.
The details of the entity query set 
is provided in the supplementary material~(Sec.~\ref{Supp: Entity Description Base and Position Set}). 
Then $Q$ will be passed into a fusion module with the image representation $\mathcal{V}$ for alignment.
The fusion module consists of multiple Transformer Decoder layers,
with $Q$ as Query, and $\mathcal{V}$ as Key and Value. The outputs are further fed into two MLPs, independently infer the existence of the entity and the entity's position:
\begin{align}
    \{ \hat{s},\text{ } \hat{p},\text{ } \hat{m} \} = \Phi_{\text{fusion}}(\mathcal{V},\text{ } Q),
\end{align}
where $\hat{s} \in \mathbb{R}^{|Q|}$ represents the existence prediction for each entity query, and $\hat{p} \in \mathbb{R}^{|Q| \times d^{\prime}}$ represents the predicted position for all entities. 
\textbf{Note that}, $\hat{m} \in \mathbb{R}^{H \times W}$ denotes the average of the cross-attention maps sourced from Transformer layers and is up-sampled to the size of input image with nearest interpolation. $\hat{m}$ is used for grounding at inference,
as it naturally acts as a segmentation heatmap. 
During training, we will not directly calculate any loss on it.

\vspace{3pt}
\noindent \textbf{Discussion.}
Adopting Transformer decoder enables to compute correspondences between entities and images at patch-level. Consequently, image features $\mathcal{V}$ are more suitable for downstream segmentation tasks and the average of cross-attention maps in each layers can be used directly for \textbf{zero-shot} grounding, providing explainable for diagnosis. Besides, the default self-attention on queries in Transformer structure can also build relationships across entities.

\subsection{Training}
\label{sec:training}
Given a set of encoded triplets $\{e, p, l\}$ extracted form the pairing reports ${\mathcal{T}}$, we can compute training loss on the output of fusion module. 
For the existence prediction $\hat{s}$, we use binary cross-entropy with the corresponding ``exist'' labels $l$, and if $l$ is $-1$ we just pass this label, denoted as $\mathcal{L}_\text{cls}$. 
To supervise the position prediction for each entity query, 
we adopt contrastive learning. We form a position set with top $|P|$ common position embeddings as a position set, $P = \{p_1, p_2, \cdots, p_{|P|} \}$, randomly sample $M$ negative position embeddings from it, and use the corresponding position embedding $p$ from triplets as positive:
\begin{equation}
    \begin{split}
        \mathcal{L}_{\text{loc}} = -\frac{1}{|Q|} \sum_{k=1}^{|Q|}\frac{e^{\langle \hat{p}_k,p_k \rangle}}{e^{\langle \hat{p}_k,p_k \rangle} + \sum_{u=1}^{M}e^{\langle \hat{p}_k,P_{\mathcal{I}(k,u)} \rangle}},
    \end{split}
\end{equation}
where $\langle \cdot , \cdot \rangle$ 
represents the inner product of two vectors and $\mathcal{I}(\cdot,\cdot)$ is a random index sampling function. The position embeddings are un-normalized in calculation. 
\textbf{Note that}, some entities may not be mentioned in the report and thus, we can not find corresponding labels in triplets. 
We simply ignore the corresponding predictions while computing loss. 

The final loss is the sum of the two:
\begin{equation}
    \mathcal{L}_{\text{total}} = \alpha_1 \mathcal{L}_{\text{loc}} 
                               + \alpha_2 \mathcal{L}_{\text{cls}},
\end{equation}
where $\alpha_1, \alpha_2$ refer to two hyper-parameters controlling the ratio of the two losses, and we set them to be $1.0$ by default. 

\vspace{5pt}
\noindent \textbf{Discussion.} 
In contrast to the existing approaches~\cite{zhang2020contrastive} that align images with entire reports, our training paradigm with triplets provides supervision at a more fine-grained entity level, rather than the global alignment between image and reports as has often done in existing approaches.

\subsection{Inference} 
\label{sec: infer}
At inference time, given a test image, 
we can directly infer the existence of certain entities/disease, 
and ground their visual evidence. 
In particular, for the entities that have appeared in the entity query set $Q$, we simply adopt the corresponding elements from $Q$, 
while for those unseen ones, we replace the entity with a brief description provided by the user, and treat it as an extra query added to entity query set $Q$, \textbf{resembling zero-shot inference}.
The existence output $\hat{s}$ can be directly applied for classification, the average cross-attention $\hat{m}$ between the target entity and the visual features are used for grounding. 

\begin{table*}[t]
\footnotesize
\centering
\setlength{\tabcolsep}{10pt}
\begin{tabular}{c|ccc|ccc|ccc}
\toprule
Dataset & \multicolumn{3}{c|}{RSNA Pneumonia} & \multicolumn{3}{c|}{SIIM-ACR Pneumothorax} & \multicolumn{3}{c}{ChestX-ray14} \\ 
Methods        & AUC$\uparrow$      & F1$\uparrow$       & ACC$\uparrow$   & AUC$\uparrow$      & F1$\uparrow$       & ACC$\uparrow$ &  AUC$\uparrow$      & F1$\uparrow$       & ACC$\uparrow$   \\ \toprule
ConVIRT~\cite{zhang2020contrastive}          & 0.8042 & 0.5842 & 0.7611 & 0.6431 & 0.4329   & 0.5700 & 0.6101 & 0.1628 & 0.7102\\
GLoRIA~\cite{huang2021gloria}           & 0.7145 & 0.4901 & 0.7129   & 0.5342 & 0.3823 & 0.4047 & 0.6610  & 0.1732 & 0.7700\\
BioViL~\cite{boecking2022making}           & 0.8280    & 0.5833   & 0.7669 & 0.7079  & 0.4855   & 0.6909  &  0.6912 & 0.1931 & 0.7916 \\ 
CheXzero~\cite{tiu2022expert} & 0.8579 & 0.6211 & 0.7942 & 0.6879 & 0.4704 & 0.5466 & 0.7296 & 0.2141 & 0.8278\\
\midrule
Ours             & \textbf{0.8694}    & \textbf{0.6342} & \textbf{0.8002} & \textbf{0.8924}    & \textbf{0.6833} & \textbf{0.8428}  & \textbf{0.7676} & \textbf{0.2525} & \textbf{0.8619}\\ \bottomrule
\end{tabular}
\vspace{-0.2cm}
\caption{Comparison with other state-of-the-art methods on zero-shot classification task.  AUC, F1 and ACC scores are reported. For ChestX-ray14, the metrics all refer to the macro average on the 14 diseases.}
\label{zero-shot_class}
\end{table*}

\begin{table*}[t]
\hspace{\fill}
\parbox[t]{.6\textwidth}{
\footnotesize
\centering
\setlength{\tabcolsep}{8pt}
\begin{threeparttable}
\begin{tabular}{c|ccc|ccc}
\toprule
Prompt Type & \multicolumn{3}{c|}{Direct Covid-19}  & \multicolumn{3}{c}{Covid-19 Description}\\
Methods     & AUC$\uparrow$   & F1$\uparrow$  & ACC$\uparrow$  & AUC$\uparrow$ & F1$\uparrow$  & ACC$\uparrow$\\ \toprule
ConVIRT~\cite{zhang2020contrastive}      & 0.6159  & 0.7057 & 0.6113 & 0.5208  & 0.6902 & 0.5266 \\
GLoRIA~\cite{huang2021gloria}        & 0.6319  & 0.6938 & 0.5710 & 0.6659  & 0.7007 & 0.6083 \\
BioViL~\cite{boecking2022making}       & 0.6137  & 0.6958 & 0.5461  & 0.5382  & 0.6910 & 0.5375 \\
CheXzero~\cite{tiu2022expert} & 0.6462 & 0.7369 & 0.6629 & 0.6667 & 0.6400 & 0.6578\\
\midrule
Ours        & 0.6561  & 0.7066 & 0.5917 & \textbf{0.7396}  & \textbf{0.7670} & \textbf{0.7006}   \\ \bottomrule
\end{tabular}
\end{threeparttable}
}
\hspace{\fill}
\parbox[t]{.38\textwidth}{
\caption{Comparison with other state-of-the-art methods on zero-shot Covid-19 classification task. AUC, F1 and ACC scores are reported. \textit{``Direct covid-19''} refers to directly use ``Covid-19'' to construct the prompt sentence while \textit{``Covid-19 Description''} refers to replace the name ``Covid-19'' with its description.  }
\label{zero-shot-class-covid-19}
}
\vspace{-0.7cm}
\end{table*}

\section{Experiment}
In this section, we start by introducing the dataset used for experiments, 
{\em e.g.}, pre-training, and various downstream datasets. 
Then we describe the implementation details and the considered baselines.

\subsection{Pre-training Dataset}
\noindent \textbf{MIMIC-CXR v2~\cite{johnson2019mimic,goldberger2000physiobank}}
consists of over 227k studies of paired image-report data, they are sourced from 65,379 patients at different scanning. Each study can have one or two images~(different scan views), totaling 377,110 images. 

\subsection{Datasets for Downstream Tasks}
\noindent \textbf{ChestX-ray14~\cite{wang2017chestx}}
contains 112,120 frontal-view X-ray images of 30,805 unique patients, 
collected from the year of 1992 to 2015 by NIH(National Institutes of Health),
with labels of 14 common diseases provided. 
We split the dataset into $0.8/0.1/0.1$ for train/valid/test. 

\vspace{3pt}
\noindent \textbf{RSNA Pneumonia~\cite{shih2019augmenting}}
contains more than 260k frontal-view chest X-rays with corresponding pneumonia opacity masks collected by RSNA (Radiological Society of North America). Commonly, it is treated as a classification tasks~\cite{huang2021gloria,boecking2022making}.
We split the dataset into $0.6/0.2/0.2$ for train/valid/test.

\vspace{3pt}
\noindent \textbf{SIIM-ACR Pneumothorax~\cite{kaggle-siim}} 
contains more than 12k frontal-view chest X-rays with pneumothorax masks collected by SIIM-ACR~(Society for Imaging Informatics in Medicine and American College of Radiology). Similarly to RSNA Pneumonia dataset, it can be both used as classification and segmentation tasks. We split the dataset into $0.6/0.2/0.2$ for train/valid/test.

\vspace{3pt}
\noindent \textbf{COVIDx CXR-2~\cite{pavlova2021covid} and COVID Rural~\cite{tang2020deep,desai2020chest}}
aim to evaluate on diagnosing COVID-19. 
COVIDx CXR-3 contains 29,986 images from 16,648 patients with COVID-19 classification labels. We use it as a classification dataset and split it into $0.7/0.2/0.1$ for train/valid/test. Additionally, we use COVID Rural dataset for COVID-19 segmentation. 
It contains more than 200 chest X-rays with segmentation masks, and we split it into $0.6/0.2/0.2$ for train/valid/test.

\vspace{3pt}
\noindent \textbf{Edema Severity~\cite{chauhan2020joint}} 
contains 6,524 examples from MIMIC-CXR with pulmonary edema severity labels (0 to 3, increasing severity) extracted from the radiology reports. Of these, 141 radiologists were examined by radiologists, and consensus was reached on severity level. It can be seen as a typical fine-grained classification task. 
We split the dataset into $0.6/0.2/0.2$ for train/valid/test.

\subsection{Implementation}
\label{ID}
This section describes the implementation for architectures. In \textbf{Pre-training}, the triplets extraction module and text encoders used in triplets encoding are all fixed,  while the visual encoder and fusion module are trained end-to-end on the image-text pairs. In \textbf{Fine-tuning}, we adopt ResNet50~\cite{he2016deep} initialized with image encoder for classification, and 
 ResUNet~\cite{diakogiannis2020resunet} initialize its encoder with our pre-trained image encoder for segmentation. More details about exact values of different parameters and training progress can be found in supplementary material~(Sec.~\ref{Supp: Details of Implementation})

\subsection{Baselines}
We compare with various existing state-of-the-art medical image-text pre-train methods, namely, ConVIRT~\cite{zhang2020contrastive}, GLoRIA~\cite{huang2021gloria}, BioViL~\cite{boecking2022making} and CheXzero~\cite{tiu2022expert}. 
Since ConVIRT and GLoRIA are pre-trained on an in-house dataset, 
we re-train their models on MIMIC-CXR dataset for fair comparison.
For BioViL, we use the officially released models by the authors. 
For zero-shot setting, we use the prompt as mentioned by BioViL~\cite{boecking2022making} and compare to the very recent method~(CheXzero~\cite{tiu2022expert}) that has shown to have better zero-shot diagnosis ability than radiologists. For fine-tuning, we all use the same setting as described in Sec.~\ref{ID}.

\subsection{Metrics}
\vspace{3pt}
\noindent \textbf{AUC, F1 and ACC} are measured for classification tasks. F1 comprehensively measures the recall and precision of the model,
and ACC is the short of Accuracy. The final binary prediction threshold is chosen to maximise the F1 score. 
The ACC score is also calculated under this threshold.

\vspace{3pt}
\noindent \textbf{Pointing Game} is used for evaluating the grounding performance. 
In specific, we extract the region with max response in the output heat-map, for one instance, if the region hit the ground-truth mask, it is considered a positive prediction, otherwise negative. Finally, accuracy can be calculated as the pointing game score. 

\vspace{3pt}
\noindent \textbf{Dice and IOU} are commonly used for segmentation tasks. 
For zero-shot segmentation, we search the segmentation threshold with 0.01 interval for all methods, and report the maximal Dice score for each model.

\vspace{3pt}
\noindent \textbf{Precision and Recall} refer to  the detection Precision and Recall. 
For medical, it is important that lesions are detected even without fine segmentation. Additionally, in some hard cases, especially for the zero-shot setting, Dice and IOU may be too strict to reflect the performance difference. Precision and recall scores can compensate for these. We choose the IOU threshold as 0.1 to calculate the scores.

\begin{table*}[t]
\footnotesize
\centering
\setlength{\tabcolsep}{5.5pt}
\begin{subtable}[t]{0.57\linewidth}
\begin{threeparttable}
\begin{tabular}{c|ccccc}
\toprule
Methods  &  Pointing Game$\uparrow$ & Recall$\uparrow$    & Precision$\uparrow$    & IoU$\uparrow$    & Dice$\uparrow$   \\ \toprule
GLoRIA~\cite{huang2021gloria}   & 0.7607 & 0.8330 & 0.1621 & 0.2182 & 0.3468 \\
BioViL~\cite{boecking2022making}    & 0.8342     & 0.8521 & 0.5034 & 0.3029 & 0.4386 \\\midrule
Ours      & \textbf{0.8721}    & \textbf{0.8661} & \textbf{0.6420} & \textbf{0.3172} & \textbf{0.4649} \\ \bottomrule
\end{tabular} 
\end{threeparttable}
\caption{Zero-shot grounding on Pneumonia}
\label{zero-shot-seg-RSNA}
\end{subtable}
\begin{subtable}[t]{0.42\linewidth}
\begin{threeparttable}

\begin{tabular}{c|ccc}
\toprule
Methods  &  Pointing Game$\uparrow$ & Recall$\uparrow$    & Precision$\uparrow$    \\ \toprule
GLoRIA~\cite{huang2021gloria}  & 0.0651     & 0.2377 & 0.0585 \\
BioViL~\cite{boecking2022making}    & 0.0252    & 0.1963 & 0.1429 \\\midrule
Ours      & \textbf{0.1975}    & \textbf{0.3562} & \textbf{0.1940}  \\ \bottomrule
\end{tabular} 
\end{threeparttable}
\caption{Zero-shot grounding on Pneumothorax}
\label{zero-shot-seg-SIIM}
\end{subtable}
\vspace{-0.2cm}
\caption{Comparison with other state-of-the-art methods on zero-shot region grounding tasks. (a) shows the results on RSNA Pneumonia dataset. (b) shows the results on SIIM-ACR Pneumothorax dataset.  The pneumothorax region tends to be thin and narrow and much more challenging for grounding, we thus only consider pointing game, recall, and precision. Our method can achieve better performance on different metrics, especially on the pointing game score. ConVIRT and CheXzero can not realize this function.}
\label{zero-shot-seg}
\end{table*}

\begin{table*}[h]
\footnotesize
\centering
\setlength{\tabcolsep}{ 4pt}
\begin{threeparttable}

\begin{tabular}{c|ccccc|ccccc}
\toprule
Prompt Type & \multicolumn{5}{c|}{Direct covid-19}  & \multicolumn{5}{c}{Covid-19 Description}\\
Methods  &  Pointing Game$\uparrow$ & Recall$\uparrow$    & Precision$\uparrow$    & IoU$\uparrow$    & Dice$\uparrow$  &  Pointing Game$\uparrow$ & AR$\uparrow$    & AP$\uparrow$    & IoU$\uparrow$    & Dice$\uparrow$ \\ \toprule
GLoRIA~\cite{huang2021gloria}   &  0.0364    & 0.2906 &  0.1073& 0.0645 & 0.1141 &   
 0.2727 & 0.2821  &  0.1336&  0.0596 & 0.1075\\
BioViL~\cite{boecking2022making}    &   0.4000   &  0.2564 &   0.2703 & 0.1198 &  0.1967 &  0.1818    & 0.2393  & 0.1637  &  0.0861 & 0.1427\\\midrule
Ours     & 0.1818 &   0.1880  & 0.1497
 & 0.0747  &   0.1289 & \textbf{0.5818}  &   \textbf{0.5214}  &  \textbf{0.4959} & \textbf{0.1373} &  \textbf{0.2278} \\ \bottomrule
\end{tabular} 
\end{threeparttable}
\vspace{-0.2cm}
\caption{ Comparison with other state-of-the-art methods on zero-shot covid-19 opacity region grounding task.  \textit{``Direct covid-19''} refers to directly use ``Covid-19'' to construct the prompt sentence for entity encoding while \textit{``Covid-19 Description''} refers to replace the name ``Covid-19`` with its description. Our method can achieve better performance on different metrics. }
\label{zero-shot-seg-covid}
\end{table*}


\begin{table*}[t]
\footnotesize
\centering
\setlength{\tabcolsep}{ 6pt}
\begin{tabular}{c|ccc|ccc|ccc|ccc}
\toprule
 Dataset        & \multicolumn{3}{c|}{Pneumonia}         & \multicolumn{3}{c|}{Pneumothorax}  & \multicolumn{3}{c|}{Covid-19} & \multicolumn{3}{c}{ChestX-ray14}\\
Data Portion  & \multicolumn{1}{c}{1\%} & \multicolumn{1}{c}{10\%} & \multicolumn{1}{c|}{100\%} & \multicolumn{1}{c}{1\%} & \multicolumn{1}{c}{10\%} & \multicolumn{1}{c|}{100\%} & \multicolumn{1}{c}{1\%} & \multicolumn{1}{c}{10\%} & \multicolumn{1}{c|}{100\%} & \multicolumn{1}{c}{1\%} & \multicolumn{1}{c}{10\%} & \multicolumn{1}{c}{100\%} \\  \toprule                  
Scratch    &0.7107   & 0.8150  & 0.8626 & 0.4347 &0.6120     & 0.6571  & 0.7861 &0.9162 & 0.9554 &0.6005  & 0.7365 &0.7924\\
ConVIRT~\cite{zhang2020contrastive}      &0.8398 &0.8562 &  0.8761  & 0.7134  &0.7826 & 0.9004   & 0.8675  &0.9541 & 0.9726 &0.6615 & 0.7658 & 0.8128\\
GLoRIA~\cite{huang2021gloria}    & 0.8599 &  0.8666   & 0.8846  &0.7439  & 0.8538 & 0.9014    &0.9065 & 0.9381  & 0.9728 &0.6710 &0.7642 & 0.8184\\
BioViL~\cite{boecking2022making}      & 0.8233&  0.8538  & 0.8836  &0.6948  &0.7775& 0.8689   & 0.8989   &0.9529 &0.9729 &0.6952 & 0.7527 & 0.8245\\\midrule
Ours       & \textbf{0.8731} &  \textbf{0.8799}  & \textbf{0.8931} & \textbf{0.8527}   & \textbf{0.9071}& \textbf{0.9188}    &\textbf{0.9224}   & \textbf{0.9657} 
& \textbf{0.9729} &\textbf{0.7721} & \textbf{0.7894} & \textbf{0.8323}\\ \bottomrule
\end{tabular}
\vspace{-0.2cm}
\caption{ Comparison of AUC scores with other state-of-the-art methods on fine-tuning classification task. The macro average of AUC scores on 14 diseases are reported for ChestX-ray14 dataset.}
\label{fine-tuning-class}
\vspace{-0.1cm}
\end{table*}

\begin{table*}[!htb]
\footnotesize
\centering
\setlength{\tabcolsep}{ 10pt}
\begin{tabular}{c|ccc|ccc|ccc}
\toprule
 Diseases                 & \multicolumn{3}{c|}{Pneumonia} & \multicolumn{3}{c|}{Pneumothorax} & \multicolumn{3}{c}{Covid-19}\\
Data Portion  & \multicolumn{1}{c}{1\%} & \multicolumn{1}{c}{10\%} & \multicolumn{1}{c|}{100\%} & \multicolumn{1}{c}{1\%} & \multicolumn{1}{c}{10\%} & \multicolumn{1}{c|}{100\%} & \multicolumn{1}{c}{1\%} & \multicolumn{1}{c}{10\%} & \multicolumn{1}{c}{100\%} \\  \toprule                  
Scratch    & 0.4347 &0.6047     & 0.7068 &0.2133   & 0.3323  & 0.7447  & 0.1481 &0.2367 & 0.3228\\
ConVIRT~\cite{zhang2020contrastive}        & 0.5706  &0.6491 & 0.7201 &0.5406 &0.6121 &  0.7352  & 0.1995  &0.2724 & 0.3737\\
GLoRIA~\cite{huang2021gloria}      &0.6555  & 0.6907 & 0.7328& 0.5673&  0.5778   & 0.7694    &0.1889 & 0.2809  & 0.3869\\
BioViL~\cite{boecking2022making}       &0.6824  &0.7038& 0.7249& 0.6267&  0.6998  & 0.7849    & 0.2113   &0.3239 &0.4162\\\midrule
Ours        & \textbf{0.7064}   & \textbf{0.7162}& \textbf{0.7579}  & \textbf{0.6659} &  \textbf{0.7210}  & \textbf{0.7937}  &\textbf{0.2445}   & \textbf{0.3539}
& \textbf{0.4399}\\ \bottomrule
\end{tabular}
\vspace{-0.2cm}
\caption{ Comparison of Dice scores with other state-of-the-art methods on fine-tuning segmentation tasks. Three diseases are reported, and for each disease, three data portions, $1\%$, $10\%$, $100\%$ are adopted to show the performance change under different data amounts.}
\label{fine-tuning-seg}
\vspace{-0.1cm}
\end{table*}

\begin{table*}[!htb]
\footnotesize
\centering
\setlength{\tabcolsep}{ 3.5pt}
\begin{tabular}{c|ccc|ccc|ccc|ccc|ccc}
\toprule
                                              \multirow{2}{*}{Methods} & \multicolumn{3}{c|}{0}                                                        & \multicolumn{3}{c|}{1}               & \multicolumn{3}{c|}{2}          & \multicolumn{3}{c|}{3}             & \multicolumn{3}{c}{AVG}        \\ 
& AUC$\uparrow$                             & F1$\uparrow$                              & ACC$\uparrow$      & AUC$\uparrow$        & F1$\uparrow$          & ACC$\uparrow$      & AUC$\uparrow$      & F1$\uparrow$       & ACC$\uparrow$      & AUC$\uparrow$      & F1$\uparrow$          & ACC$\uparrow$      & AUC$\uparrow$      & F1$\uparrow$       & ACC$\uparrow$      \\ \midrule
Scratch                                       & 0.7631                        & 0.7036                        & 0.6738 & 0.5383 & 0.3593 & 0.3223 & 0.6692 & 0.4328 & 0.7012 & 0.8420 & 0.5694 & 0.8770 & 0.7031 & 0.5163 & 0.6436 \\
ConVIRT~\cite{zhang2020contrastive}                                       &  0.8453 & \textbf{0.7769} & \textbf{0.7793} & 0.6099 & 0.3938  & 0.4629 & 0.7202 & 0.4843  & 0.6445 & 0.9047 & 0.6154 & 0.8809 & 0.7700 & 0.5676 & 0.6919 \\
GLoRIA~\cite{huang2021gloria}                                        & 0.8304 & 0.7577 & 0.7520 & 0.6208 & 0.3991 & 0.4922 & 0.7339 & 0.4958 & \textbf{0.7037} & \textbf{0.9246} & \textbf{0.6667} & 0.9102 & 0.7774 & 0.5798 & 0.7145 \\
BioViL~\cite{boecking2022making}                                        & 0.8034                        & 0.7378                        & 0.7148 & 0.6035 & 0.3912 & 0.4570 & 0.6860 & 0.4497 & 0.6777 & 0.9229  & 0.6500        & 0.9160 & 0.7540 & 0.5572 & 0.6914 \\\midrule
Ours                                          & \textbf{0.8502}                        & 0.7646                       & 0.7539 & \textbf{0.6641} & \textbf{0.4140} & \textbf{0.5392} & \textbf{0.7605} & \textbf{0.5266} & 0.7031 & 0.8845 & 0.6250       & \textbf{0.9160} & \textbf{0.7898} & \textbf{0.5826} & \textbf{0.7280} \\ \bottomrule
\end{tabular}
\vspace{-0.2cm}
\caption{ Comparison with other state-of-the-art methods on fine-tuning edema severity grading multi-class classification task. AUC score is reported in the Table. ``0,1,2,3'' in the table represents the severity level and final average scores are reported.}
\vspace{-0.1cm}
\label{fine-tuning-grading}
\end{table*}

\section{Results}
In this section, we will report the experimental results. 
In general, we split the results into two parts: zero-shot setting and fine-tuning setting. In the zero-shot case~(Sec.~\ref{Zero-shot}), we carry out the ablation study and compare it with the other SOTA image-text pre-train methods. 
We mainly consider classification and segmentation tasks;
In the fine-tuning case~(Sec.~\ref{Finetuning}), we evaluate the model's transferability by fine-tuning the model with $1\%$, $10\%$, and $100\%$ data portion. Additionally, we also add a disease grading downstream task, which can be seen as a fine-grade classification task, showing that our pre-trained model can be transferred to the downstream tasks at ease.

\subsection{Zero-shot Evaluation}
\label{Zero-shot}
In this section, we compare our method with other state-of-the-art methods under zero-shot setting, on classification and grounding. 
Due to the space limitation, we include the entire ablation study in the supplementary material~(Sec.~\ref{Supp: Ablation}), referring to it for more details and analysis, and all comparisons here are made using our best model with position contrastive loss and entity description encoder.

\subsubsection{Classification}
\noindent \textbf{Seen Diseases.} 
As shown in Tab.~\ref{zero-shot_class},
we compare with existing methods on three widely-used datasets, demonstrating consistent performance improvement.
Specifically, on pneumonia and pneumothorax datasets, 
despite the images being collected by different clinics with different diseases, 
our model improves the AUC score from $0.83$ to $0.87$ on RSNA pneumonia dataset and from $0.71$ to $0.89$ on SIIM-ACR pneumothorax dataset, as shown in Tab.~\ref{zero-shot_class}. This shows that our method can better deal with the multi-center and multi-disease data distribution in medical. While on ChestX-ray14 dataset, we improve the average AUC scores from $0.69$ to $0.77$, we refer the reader to supplementary material~(Sec.~\ref{Sec: Detailed results on ChestX-ray14}) for detailed comparison of 14 diseases.

\vspace{3pt}
\noindent \textbf{Unseen Diseases.}
Here, we are considering a strict setting for openset classification,
in particular, we use covid-19 to evaluate the systems. 
Covid-19 is a new disease that only appeared in 2019, 
MIMIC-CXR reports collected in the year 2015 do not include any entity of covid-19, thus it requires the system to have the ability to diagnose truly unseen diseases. As shown in Tab.~\ref{zero-shot-class-covid-19}, existing approaches that only rely on disease name struggles to make the correct diagnosis. While, our proposed approach, after introducing medical knowledge, 
{\em i.e.}, using entity descriptions, can understand the complex medical entity descriptions unseen in the training set,
and significantly boost the performance from $0.66$ to $0.74$ on AUC and from $0.59$ to $0.70$ on ACC, demonstrating entity translation is vital for unseen diseases.


\subsubsection{Grounding}
In addition to the plain diagnosis, explainability can be equally critical in healthcare, improving the reliability and trustiness of the machine learning systems. Here, we consider providing explainability by grounding the abnormality in the prediction and compare against the existing approaches. Similarly, we split the diseases into seen and unseen ones, depending on whether their names have appeared in the medical reports. Specifically, ``Pneumonia'' and ``Pneumothorax'' are treated as seen, and ``Covid-19'' is treated as unseen.
 Due to the space limitation, we include visualization results in supplementary material~(Sec.~\ref{Visualization}).

\vspace{3pt}
\noindent \textbf{Seen Diseases.} 
We show the results for grounding on RSNA Pneumonia opacity and SIIM-ACR Pneumothorax collapse in Tab.~\ref{zero-shot-seg}. As shown in Tab.~\ref{zero-shot-seg-RSNA}, our proposed model surpasses existing approaches on all metrics, for example, we improve the pointing game score from $0.83$ to $0.87$, the detection Recall from $0.85$ to $0.87$, the detection precision from $0.50$ to $0.64$, the IOU from $0.30$ to $0.32$ and the Dice from $0.44$ to $0.46$. While on SIIM-ACR dataset~(Tab.~\ref{zero-shot-seg-SIIM}),
the pneumothorax region tends to be thin and narrow, 
localizing it can often be more challenging than that of opacity grounding~\cite{boecking2022making}, we thus only consider pointing game, recall, and precision. Similarly, our method can achieve significantly better performance than prior approaches.

\vspace{3pt}
\noindent \textbf{Unseen Diseases.} 
We also conduct the zero-shot grounding experiment on the unseen disease, namely, Covid-19, as shown in Tab.~\ref{zero-shot-seg-covid}. 
Our model has shown consistent improvements in all metrics, 
{\em e.g.}, boosting the pointing game score from $0.40$ to $0.58$. 
One observation to be noticed is that, results in Tab.~\ref{zero-shot-seg-covid} are mostly consistent with those in Tab.~\ref{zero-shot-class-covid-19}, 
{\em i.e.}, better classification results tend to lead to better grounding. 
Overall, our model with knowledge-enhanced language encoding has facilitated the visual encoder to learn underlying evidence relating to the diseases, therefore, leading to more interpretable representations than prior approaches.


\subsection{Fine-tuning Evaluation}
\label{Finetuning}

In this section, we consider the fine-tuning scenario, 
with the pre-trained model as initialization, and trained end-to-end on the downstream tasks. We test on three different tasks, namely, classification, segmentation, and grading. In classification and segmentation, the test splits and metrics are the same as in the ``zero-shot'' section. Grading is a new task we introduce in fine-tuning setting, which can be seen as a fine-grained classification task.

\vspace{-0.2cm}
\subsubsection{Classification}
We experiment on four different datasets, using $1\%$, $10\%$, $100\%$ of the data for fine-tuning, that is consistent with the existing work~\cite{zhang2020contrastive,huang2021gloria,boecking2022making}.
As shown in Tab.~\ref{fine-tuning-class}, our model has demonstrated substantial improvements in the AUC scores over the existing approaches across all datasets, 
reflecting that our pre-trained representation is of higher quality than existing models. We refer the readers to the supplementary material~(Sec.~\ref{Sec: Detailed results on ChestX-ray14}) for more detailed comparison results.

\vspace{-0.2cm}
\subsubsection{Segmentation}
In Tab.~\ref{fine-tuning-seg}, we conduct fine-tuning experiments on three different diseases for segmentation. 
We pick $1\%$, $10\%$, $100\%$ of the data for fine-tuning.
For all three different diseases with different image distributions, our methods surpass the existing state-of-the-art methods by a large margin,
especially under the low-data regime. 

\vspace{-0.2cm}
\subsubsection{Grading}
Besides diagnosis, grading the disease severity level also plays an important role. Here, we adopt our pre-trained features and train them for the multi-class classification task, with $0$ to $3$ representing different severity levels. As shown in Tab.~\ref{fine-tuning-grading}, for each level, the AUC, F1, and ACC scores are calculated as one class against all other ones, for example, $0$ vs $\left\{1,2,3\right\}$. Final macro average scores of four levels are computed. On the majority of severity levels, our method can achieve the best results.

\section{Conclusion}
    In this paper, we introduce a novel medical knowledge enhanced VLP model. 
    {\em First}, we propose a triplet extraction module to extract useful medical-related triplets as more useful supervision signals, simplifying complex raw reports with minimal information loss. 
    {\em Second}, we translate the entities in extracted triplets into detailed medical descriptions and embed them with a text encoder enabling the network to understand complex medical expert-level knowledge. 
    {\em Finally}, a transformer-based structure is proposed to do local region alignment. 
    In experiments, We evaluate our method on different datasets under various settings. Our method shows strong zero-shot classification and grounding abilities, even facing unseen diseases. 
    {\em Additionally}, with fine-tuning, our method still outperforms state-of-the-art methods significantly, showing the superiority of our method.

{\small
\bibliographystyle{ieee_fullname}
\bibliography{egbib}
}

\clearpage
\onecolumn
\appendix
\renewcommand*\contentsname{Supplementary}
\newpage
\section{The Entity Description Base and Position Set}
\label{Supp: Entity Description Base and Position Set}
We leverage the rad-graph NER extraction results provided by ~\cite{yu2022anatomy} and further add extra descriptions. Tab.~\ref{Table: Description Base} shows the descriptions we used to translate different entities. We have kept $75$ entities in entity query set $Q$, following~\cite{yu2022anatomy}, which covers $90\%$ entities in reports. ``Tail\_abnorm\_obs'' entity represents some tailed entities and ``exluded\_obs'' represents some entities useless for diagnosis. The last ``covid-19'' row is only referred for inference since it never appear in pre-train reports.
{
\footnotesize
\begin{longtable}{l|p{15cm}}
  \caption{The Entity description used for translate single entity name. The description can be easily found from website.}
 \label{Table: Description Base}\\
    \toprule
    Entity              & Description   \\ \toprule
    \endfirsthead
    
   \toprule
    Entity              & Description   \\ \toprule
    \endhead
    
    \bottomrule
    \endfoot
     normal             & It means the absence of diseases and infirmity, indicating the   structure is normal.\\
 clear             & The lungs are clear and normal. No evidence for other diseases   on lung.   \\
 sharp             & This means that an anatomical structure s boundary or edge is   clear and normal, meaning it is free of diseases.  \\
 sharply           & ``Sharply seen   means that an anatomical structure is clearly   visible. \\
unremarkable      & This represents some anatomical structures are normal, usually   modifying cardiac and mediastinal silhouettes.  \\
intact           & The bonny structure is complete and normal, meaning no   fractures.\\
 stable            & The modified anatomical structures are normal and stable. No   evidence for diseases. \\
 free              & It usually refers to free air and is associate with   pneumothorax, atelectasis, pneumoperitoneum and emphysema.  \\
 effusion          & A pleural effusion is accumulation of excessive   fluid in the pleural space, the potential space that surrounds   each lung. A pleural effusion infiltrates the space between the visceral   pleura and the parietal pleura.\\
 opacity           & It is defined as an area of hazy opacification due to air   displacement by fluid, airway collapse, fibrosis, or a neoplastic process. It   is causes include infections, interstitial lung disease, and pulmonary edema. \\
 pneumothorax      & A pneumothorax is an abnormal collection of air in   the pleural space between the lung and the chest   wall. It may be caused by pneumonia or fibrosis and other diseases.\\
 edema             & Pulmonary edema, also known as pulmonary congestion, is   excessive liquid accumulation in the tissue and air   spaces of the lungs. It will show fluid in the alveolar walls.\\
 atelectasis       & It is the collapse or closure of a lung resulting in reduced   or absent gas exchange. Findings can include lung opacification and loss of   lung volume. \\
 tube              & It is a surgical drain that is inserted through   the chest wall and into the pleural space or   the mediastinum to remove undesired substances such as air.\\
 consolidation     & It is a region of normally compressible lung   tissue that has filled with liquid instead of air. Consolidation must be   present to diagnose pneumonia: the signs of lobar pneumonia are   characteristic and clinically referred to as consolidation.                                                                 \\
 process           & Acute process means there is abnormality in the anotomy   structure.   \\
 abnormality       & It means the exist of diseases and infirmity, indicating the   structure is abnormal.                                                                                                           \\
 enlarge           & It usually modifies cardiac silhouette and heart. Cardiomegaly   is a medical condition in which the heart is enlarged.                                                                         \\
 tip               & It refers to the top head of the tube.                                  \\
 low               & The presence of low lung volumes may be a sign of a   restrictive lung condition such as pulmonary fibrosis or sarcoidosis.                                                                     \\
pneumonia          & Pneumonia is an inflammatory condition of   the lung primarily small air sacs known as alveoli.  Pneumonia may present with opacities. Complications such as pleural effusion may also be found increasing the diagnostic accuracy of lung consolidation and pleural effusion                                       \\
 line              & It refers to venous access line ot PICC lines.                          \\
 congestion        & Pulmonary congestion is defined as accumulation of fluid   in the lungs, resulting in impaired gas exchange and arterial hypoxemia.                                                             \\
 catheter          & catheter is a tube placed in the body to drain and   collect urine from the bladder                                                                                                             \\
 cardiomegaly      & Cardiomegaly (sometimes megacardia or megalocardia)   is a medical condition in which the heart is enlarged.                                                                                    \\
 fracture          & Fracture is a break in a rib bone.                                      \\
 air               & It refers to the free air or gas in pleural space, indicating   pneumothorax. Air displacement by fluid may lead to opacity.                                                                    \\
 tortuous          & The Aorta is slightly tortuous. Sometimes it may refer to   varicose veins                                                                                                                      \\
 lead              & It refers to the leading head of the tube.                              \\
 disease           & It means the exist of diseases and abnormalty, indicating the   structure is abnormal.                                                                                                          \\
 calcification     & Pulmonary calcification is a common asymptomatic finding.   Pulmonary calcifications are caused mainly by two mechanisms: the dystrophic   form and the metastatic form                         \\
 prominence        & It means the exist of some observation.                                 \\
 device            & It refer to some equipments like picc tub, valve catheter,   pacemaker hardware, arthroplastmarker icd defib, device support equipment and   mediport                                          \\
 engorgement       & Pulmonary vascular engorgement means obstruction of the normal   flux of blood within the blood vessel network of the lung resulting in   engorgement of pulmonary vessels                      \\
 picc              & A peripherally inserted central catheter (PICC), also called a   PICC line, is a long, thin tube that s inserted through a vein in your arm   and passed through to the larger veins near your heart.                                                                                                                   \\
 clip              & Surgical clips or vascular clips usually represent   the one kind of medical equipments.                                                                                                        \\
 elevation         & If tissues or anatomical   structures are elevated, they are raised up higher than   the normal location.                                                                                       \\
 expand            & It means the lungs are normally expanded and clear, indicating   the absence of pneumothorax.                                                                                                   \\
 nodule            & A lung nodule or pulmonary nodule is a relatively small focal   density in the lung. it may be confused with the projection of a structure of   the chest wall or skin, such as a nipple, a healing rib fracture or  lung cancer.                                                                                       \\
 wire              & Sternotomis wires means the center line of the chest.                   \\
 fluid             & It refers to the water of liquid in the lung and it may   indicate edema and other diseases.                    
 \\
  degenerative      & Degenerative disease is the result of a continuous   process based on degenerative cell changes                                                                                                 \\
  pacemaker         & pacemaker device usually represents the one kind of medical equipments.                                                                                                                       \\
 thicken           & Pleural thickening is an increase in the bulkiness of one   or both of the pulmonary pleurae. It may cause by pulmonary Infection,   empyema, tuberculosis or lung cancer.                      \\
 marking           & It represents interstitial markings or bronchovascular   markings       \\
 scar              & A scar (or scar tissue) is an area   of fibrous tissue that replaces normal tissues after   an injury.                                                                                          \\
 hyperinflate      & Hyperinflated lungs are larger-than-normal lungs as a   result of trapped air.                                                                                                                  \\
 blunt             & Blunting of the costophrenic angles is usually caused by a   pleural effusion, as already discussed. Other causes of costophrenic angle   blunting include lung disease in the region of the costophrenic angle, and   lung hyperexpansion. \\
 loss              & The etiology of lung volume loss can be listed as follow: airway obstruction or compression, obesity, scoliosis, restrictive diseases   such as pulmonary fibrosis and interstitial lung disease, tuberculosis. \\
 widen             & The mediastinum is not widened or enlarged. \\
 collapse          & Collapse lung refers to pneumothorax or atelectasis.\\
 density           & The density (more precisely, the volumetric   mass density; also known as specific mass), of a substance is   its mass per unit volume.\\
 emphysema         & Emphysema, or pulmonary emphysema, is a lower   respiratory tract disease, characterized by air-filled spaces   (pneumatosis) in the lungs, that can vary in size and may be very large.\\
 aerate            & Aeration (also   called aerification or aeriation) is the process by   which air is circulated through, mixed with   or dissolved in a liquid or other substances that act as   a fluid (such as soil). \\
 mass              & A lung mass is an abnormal growth or area in the lungs   and it can also view as lung cancer.  \\
 crowd             & Crowding of the bronchovascular structures is an   important direct sign of volume loss. The atelectatic lung enhances densely   after contrast administration because of closeness of the pulmonary arteries   and arterioles within the collapsed lobe.  \\
 infiltrate        & A pulmonary infiltrate is a substance denser   than air, such as pus, blood, or protein, which lingers   within the parenchyma of the lungs. Pulmonary infiltrates are   associated with pneumonia, tuberculosis and sarcoidosis. \\
 obscure           & Some anatomy structures are not clear and is   difficult to understand or see. \\
 deformity         & It means some body parts are abnormal or unjuried.  \\
 hernia            & Lung hernia (Sibson hernia) is a protrusion of lung outside of   thoracic wall. the hernia is noted after chest trauma, thoracic surgery or   certain pulmonary diseases.\\
 drainage          & Tube drainage represents the one kind of medical equipment.\\
 distention        & Distension generally refers to an enlargement, dilation, or   ballooning effect. It may refer to: Abdominal distension. \\
 shift             & The mediastinal shift is the deviation of the mediastinal   structures towards one side of the chest cavity, usually seen on chest   radiograph. It indicates a severe asymmetry of intrathoracic pressures. \\
 stent             & Tracheal stent represents the one kind of medical equipments            \\ 
 pressure          & Pulmonary venous pressure is intermediate between mean   PAP and LAP over all physiologic pressures  \\
 lesion            & Lung nodules, pulmonary nodules, white spots, lesions—these   terms all describe the same phenomenon: an abnormality in the lungs. \\
 finding           & Some observation on body parts, usually indicating abnormalty.  \\
 borderline        & Borderline size of the cardiac silhouette means the cardiac   silhouette is not enlarged and normal. \\
 hardware          & It represents the one kind of medical equipments.   \\
 dilation          & The state of being larger or more open than normal  \\
 chf               & Heart failure — sometimes known as congestive heart failure —   occurs when the heart muscle doesn't pump blood as well as it should. When   this happens, blood often backs up and fluid can build up in the lungs,   causing shortness of breath.  \\
 redistribution    & If the pulmonary edema is due to heart failure or fluid   overload, you may also see cardiomegaly and distension of the pulmonary   veins, particularly in the upper lung fields.\\
 aspiration        & Aspiration pneumonia occurs when food or liquid is breathed   into the airways or lungs, instead of being swallowed.   \\
 tail\_abnorm\_obs & Some very rare diseases.\\
 excluded\_obs     & Some meaningless observations.  \\\midrule
 covid-19            & It is a contagious disease caused by a virus. Ground-glass   opacities,  consolidation, thickening,   pleural effusions commonly appear in infection.  \\

\end{longtable}  
}

 Additionally, we keep $51$ positive positions, following~\cite{yu2022anatomy}, to form the position set $P$, as \textit{\{trachea, left\_hilar, right\_hilar, hilar\_unspec, left\_pleural, right\_pleural,	pleural\_unspec, heart\_size, heart\_border, left\_diaphragm, right\_diaphragm,	diaphragm\_unspec, retrocardiac, lower\_left\_lobe, upper\_left\_lobe, lower\_right\_lobe	middle\_right\_lobe, upper\_right\_lobe, left\_lower\_lung, left\_mid\_lung, left\_upper\_lung	left\_apical\_lung, left\_lung\_unspec, right\_lower\_lung, right\_mid\_lung, right\_upper\_lung	right\_apical\_lung, right\_lung\_unspec, lung\_apices, lung\_bases, left\_costophrenic	right\_costophrenic, costophrenic\_unspec, cardiophrenic\_sulcus, mediastinal, spine	clavicle, rib, stomach, right\_atrium, right\_ventricle, aorta, svc, interstitium, parenchymal, cavoatrial\_junction, cardiopulmonary, pulmonary, lung\_volumes, unspecified, other\}.} ``Other'' is used to reprepresnt some tailed positions.

\section{Implementation Details}
\label{Supp: Details of Implementation}

\vspace{2pt}
\noindent \textbf{Model architecture.} 
As input to the model, images are resized into $224\times 224 \times 3$. 
We use the first four layers of ResNet50~\cite{he2016deep} as our visual backbone~($\Phi_\text{visual}$), and adopt a MLP layer to transform the output feature dimension into $d = 256$. As a result, the  output feature maps from visual encoder is $\mathcal{V} \in \mathbb{R}^{14 \times 14 \times 256}$. 
On the report side, we extract the triplets with a pre-trained NER module, as described in~\cite{jain2021radgraph}, and compute the entity and position embedding with a pre-trained ClinicalBERT~\cite{alsentzer-etal-2019-publicly}, its default embedding dim is $d^{\prime} = 768$. We obtain $|Q| = 75$ entities and $|P| = 51 $ positions that most frequently appear in the reports, following ~\cite{yu2022anatomy}. We sample $M=7$ negative positions for each entity to calculate contrastive loss. 
In the fusion module, We adopt 4 Transformer Decoder layers with $4$ heads. 

\vspace{2pt}
\noindent \textbf{Pre-training.} 
At this stage, both the pre-process operation and language encoding use pre-trained networks, while the visual encoder and fusion module are trained end-to-end. We use AdamW~\cite{loshchilov2017decoupled} optimizer with $lr =1\times 10^{-4}$ and $lr_ \text{warm up} = 1\times 10^{-5}$. We train on a GeForce RTX 3090 GPU with batch size 32 for 60 epochs. First $5$ epochs are set for warming up.


\vspace{2pt}
\noindent \textbf{Fine-tuning.} 
For the downstream tasks, with large amount of training data, we can fine-tune the model end-to-end, with our pre-trained visual backbone as initialization. Specifically, for image classification task, we adopt ResNet50~\cite{he2016deep} and initialize its first four layers with our pre-trained visual encoder. For image segmentation task, we use ResUNet~\cite{diakogiannis2020resunet} as backbone and initialize its encoder with our pre-trained image encoder.

\section{Ablation Study}
\label{Supp: Ablation}
Our final method mainly contains three key parts, transformer-based entity-query fusion module, position location contrastive loss~(PosCL), and entity translation~(ET) encoding. We gradually remove the modules to analyze their effectiveness. ``w/o~(ET)'' refers to removing ET module and ``w/o~(PosCL + ET)'' refers to only maintaining the fusion module with basic CE loss.  We cannot dismiss the transformer-based entity-query fusion module as it is the most basic module to support our pre-training. Tab.~\ref{Ablation-zero-shot_class} and Tab.~\ref{Ablation-zero-shot-seg} shows the quantitative results. 

\vspace{2pt}
\noindent \textbf{Entity-query Fusion module.}
The lines about ``w/o~(PosCL + ET)'' in tables demonstrate the performance of the basic model modified only by base entity existence CE loss. This model can exceed many former methods. This proves our assumption that the complex syntax will hurt the network to capture the useful entities significantly and our pre-process operation and entity-level supervision can greatly relieve the problem.

\vspace{2pt}
\noindent \textbf{Position Contrastive Loss.}
The PosCL can significantly help the network to ground the abnormalities. As shown in the results by adding PosCL the classification results can be further improved, e.g., from $0.75$  to $0.76$ on AUC in ChestX-ray14 dataset. Besides classification, location contrastive loss can bring more gain in grounding. These results show position is another vital element in reports especially for grounding tasks. Our extracted triplets can conclude and clean the reports with little information loss and make the network learn the report information more straightforward.

\vspace{2pt}
\noindent \textbf{Entity Translation.}
By adding entity translation, we want to realize two goals. \textit{First}, in addition to just learning from the image-report data, the network can actively learn the relationship between different entities based on the entity descriptions. As shown in tables, adding descriptions in most scenarios can help the network better understand the entity and bring gain to the final metric scores. \textit{Second}, 
more importantly, the entity translation enables our model to \textbf{handle openset new diseases}. If excluding TE  and prompting the entities only with their names as former works, the performance of our method will drop significant when facing unseen diseases which is discussed in zero-shot classification for Covid-19 at main body.

\begin{table*}[!htb]
\footnotesize
\centering
\setlength{\tabcolsep}{10pt}
\begin{tabular}{c|ccc|ccc|ccc}
\toprule
Dataset & \multicolumn{3}{c|}{RSNA Pneumonia} & \multicolumn{3}{c|}{SIIM-ACR Pneumothorax} & \multicolumn{3}{c}{ChestX-ray14} \\ 
Methods        & AUC$\uparrow$      & F1$\uparrow$       & ACC$\uparrow$   & AUC$\uparrow$      & F1$\uparrow$       & ACC$\uparrow$ &  AUC$\uparrow$      & F1$\uparrow$       & ACC$\uparrow$   \\ \toprule

 w/o~(PosCL + ET)    & 0.8532     & 0.6079   & 0.7669  & 0.8768     & 0.6672   & 0.8187 & 0.7502 & 0.2374 & 0.8541\\
 w/o~(ET)          & 0.8537     & 0.6241   & \textbf{0.8146} & \textbf{0.9017}     & \textbf{0.7008}   & \textbf{0.8584}& 0.7621 & 0.2452 & 0.8606\\
Ours             & \textbf{0.8694}    & \textbf{0.6342} & 0.8002 & 0.8924    & 0.6833 & 0.8428  & \textbf{0.7676} & \textbf{0.2525} & \textbf{0.8619}\\ \bottomrule
\end{tabular}

\caption{Ablation study on zero-shot classification task.  AUC, F1 and ACC scores are reported. For ChestX-ray 14, the metrics all refer to the macro average on the 14 diseases.}

\label{Ablation-zero-shot_class}
\end{table*}

\begin{table*}[!htb]
\footnotesize
\centering
\setlength{\tabcolsep}{5.6pt}
\begin{subtable}[t]{0.57\linewidth}
\begin{threeparttable}
\begin{tabular}{c|ccccc}
\toprule
Methods  &  Pointing Game$\uparrow$ & Recall$\uparrow$    & Precision$\uparrow$    & IoU$\uparrow$    & Dice$\uparrow$   \\ \toprule
w/o~(PosCL + ET)    &0.7979 & \textbf{0.8961} & 0.4036 & 0.2783 & 0.4230   \\ 
w/o~(ET)   & 0.8424 & 0.8226 & \textbf{0.6520}& 0.3118 & 0.4610   \\ 
Ours      & \textbf{0.8721}    & 0.8661 & 0.6420 & \textbf{0.3172} & \textbf{0.4649} \\ \bottomrule
\end{tabular} 
\end{threeparttable}
\caption{Zero-shot grounding on Pneumonia}
\end{subtable}
\begin{subtable}[t]{0.42\linewidth}
\setlength{\tabcolsep}{4.6pt}
\begin{threeparttable}

\begin{tabular}{c|ccc}
\toprule
Methods  &  Pointing Game$\uparrow$ & Recall$\uparrow$    & Precision$\uparrow$    \\ \toprule
w/o~(PosCL + ET)    &0.1786 & 0.3151 & 0.1336   \\ 
w/o~(ET)   & \textbf{0.2080} & 0.3178  & 0.1711    \\ 
Ours      & 0.1975    & \textbf{0.3562} & \textbf{0.1940}  \\ \bottomrule
\end{tabular} 
\end{threeparttable}
\caption{Zero-shot grounding on Pneumothorax}
\end{subtable}

\caption{Ablation study on zero-shot grounding tasks. (a) shows the results on RSNA Pneumonia dataset. (b) shows the results on SIIM-ACR Pneumothorax dataset.}
\label{Ablation-zero-shot-seg}
\end{table*}

 \clearpage
\section{Detailed results on ChestX-ray14}
\label{Sec: Detailed results on ChestX-ray14}
\begin{figure}[!htb]
\hspace{\fill}
\parbox[c]{.6\textwidth}{
     \centering
    \includegraphics[width=.6\textwidth]{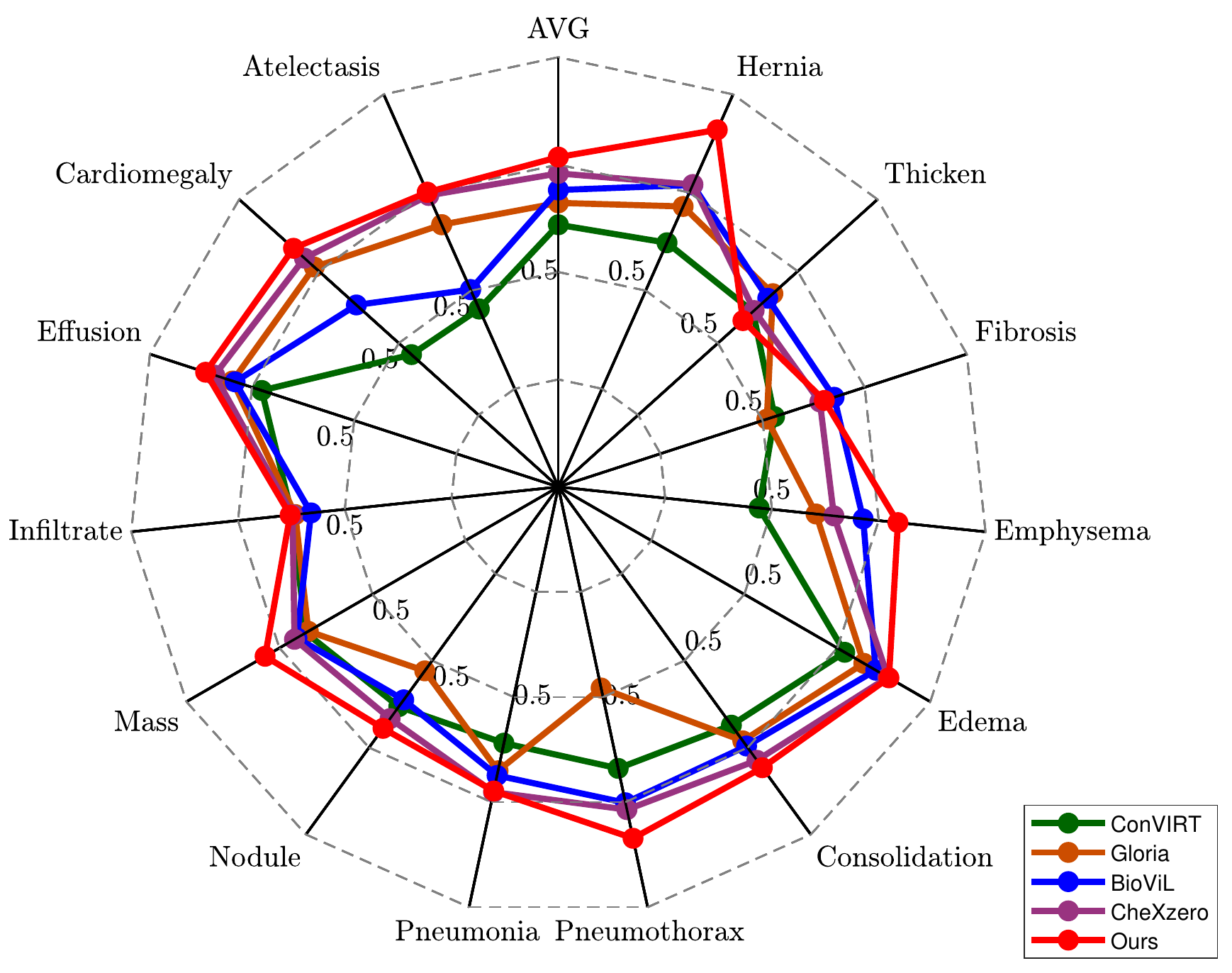}
}
\hspace{\fill}
\parbox[c]{.38\textwidth}{
\caption{The radar figure of our method and other methods of  ChestX-ray14 14 diseases. AUC scores are reported and, as shown, our method exceeds the previous state-of-the-art on most diseases.}
\label{radar-zero-shot-chestX-ray14}
}

\end{figure}

We further show the detailed performance of 14 different diseases on  ChestX-ray14 dataset. Tab.~\ref{zero-shot_class_chest-x-ray} shows the results on the zero-shot setting. Our method can exceed the former methods for most diseases. The radar Fig.~\ref{radar-zero-shot-chestX-ray14}Y shows more visually how our model compares with other solutions under the zero-shot setting. Our method can exceed the former methods for most diseases. Under $100\%$ fine-tuning settings, we achieved similarly excellent results as shown in Tab.~\ref{Detail-fine-tuing-class-chestx-ray14}.

\begin{table}[!htb]
\footnotesize
\centering
\setlength{\tabcolsep}{3pt}
\begin{tabular}{c|cccccccccccccc|c}
\toprule
Methods                & Ate.       & Car.      & Eff.          & Inf.       & Mas.                 & Nod.            & Pna.         & Pnx.      & Con.     & Ede.             & Emp.            & Fib.       & Thi.          & Her.            & AVG           \\ \toprule
ConVIRT~\cite{zhang2020contrastive}                          & 0.4533          & 0.4601         & 0.7262          & 0.6238          & 0.6790          & 0.6322   & 0.6097          & 0.6698          & 0.6855         & 0.7699          & 0.4701          & 0.5293          & 0.6098          & 0.6220             & 0.6101          \\
GLoRIA~\cite{huang2021gloria}                           & 0.6680    &  0.7647    &  0.7975    &  0.6159    & 0.6722          & 0.5293          & 0.6755          & 0.4785          & 0.7306          & 0.8212         & 0.6033          & 0.5104          &  \textbf{0.6721}   & 0.7144          & 0.6610         \\
BioViL~\cite{boecking2022making}                           & 0.5026            & 0.6328            & 0.7914            & 0.5791              &  0.7029         & 0.6126            &  0.6866     &  0.7516      & 0.7455      &  0.8533      & 0.7136   & 0.6751 & 0.6560          & 0.7692     &  0.6909    \\
CheXzero~\cite{tiu2022expert} & 0.7426 & 0.7956 & 0.8415 & 0.6223 & 0.7095 & 0.6666 & 0.7263 & 0.7679 & 0.7866 & 0.8862 & 0.6451 & 0.6402 & 0.6134 & 0.7704 & 0.7296
\\
\midrule
w/o~(PosCL + ET)  &0.7131 & 0.8100 & 0.8635 & \textbf{0.6361} & 0.7776 & 0.6740 & 0.6903 & 0.8124 & 0.7915 & 0.8869 & 0.7480 & \textbf{0.6780} & 0.6429 & 0.7784 & 0.7502\\ 
w/o~(ET) & 0.7420 & 0.8270 & \textbf{0.8663} & 0.6336 & 0.7867 & \textbf{0.6974} & \textbf{0.7238} & 0.8310 & 0.8037 & 0.8887 & 0.7865 & 0.6715 & 0.5414 & 0.8691 & 0.7621\\
ours & \textbf{0.7506} & \textbf{0.8299} & 0.8636 & 0.6280 & \textbf{0.7885} & 0.6947 & 0.7236 & \textbf{0.8361} & \textbf{0.8079} &\textbf{0.8888} & \textbf{0.7950} & 0.6511    & 0.5783 & \textbf{0.9097} & \textbf{0.7676} \\ \bottomrule

\end{tabular}
\caption{ Comparison with other state-of-the-art methods on zero-shot ChestX-ray 14 diseases classification task. For each disease, AUC score is reported and the macro average AUC score is also reported. We use the first three letters to represent one disease but for ``pneumonia''  and ``pneumothorax''  we use the first two and the last letters.
}
\label{zero-shot_class_chest-x-ray}
\end{table}

\begin{table}[!htb]
\footnotesize
\centering
\setlength{\tabcolsep}{ 3pt}
\begin{tabular}{c|cccccccccccccc|c}
\toprule
Methods & Ate. & Car. & Eff. & Inf. & Mas.        & Nod.   & Pna. & Pnx. & Con. & Ede.   & Emp.   & Fib. & Thi. & Her.  & AVG  \\ \toprule
Scratch      & 0.7835    & 0.8116     & 0.8563 & 0.6537   & 0.7788 & 0.6912 & 0.7004  & 0.8561     & 0.8090      & 0.8869  & 0.8564 & 0.7534 & 0.7454    & 0.9106 & 0.7924 \\
ConVIRT~\cite{zhang2020contrastive}      & 0.8012    & 0.8360     & 0.8511 & 0.6613   & 0.8004 & 0.7490 & 0.6998  & 0.8666     & 0.8079      & 0.9023 & 0.9014 & 0.7933 & 0.7468   & 0.9627 & 0.8128 \\
GLoRIA~\cite{huang2021gloria}       & 0.8263    & 0.8326     & 0.8596  & \textbf{0.6641}   & 0.8179 & 0.7348 & 0.7104  & 0.8452     & 0.8129     & 0.8977 & \textbf{0.9310} & 0.7886 & 0.7608   & 0.9750 & 0.8184 \\
BioViL~\cite{boecking2022making}       & 0.8185    & 0.8543     & 0.8607 & 0.6660   & 0.8302 & 0.7633 & 0.7090  & 0.8595     & \textbf{0.8287}      & 0.9031 & 0.9251 & 0.7912 & 0.7638   & 0.9696  & 0.8245  \\ \midrule
ours         & \textbf{0.8291}    & \textbf{0.8594}     & \textbf{0.8719} & 0.6565   & \textbf{0.8382} & \textbf{0.7647} & \textbf{0.7378}  & \textbf{0.8807}     & 0.8275      & \textbf{0.9083} & 0.9224 & \textbf{0.7977} & \textbf{0.7784}   & \textbf{0.9796} & \textbf{0.8323} \\ \bottomrule
\end{tabular}
\caption{ Comparison with other state-of-the-art methods on fine-tuning ChestX-ray 14 diseases classification task. For each disease, AUC score is reported and the macro average AUC score is also reported. We use the first three letters to represent one disease but for ``pneumonia''  and ``pneumothorax''  we use the first two and the last letters.}
\label{Detail-fine-tuing-class-chestx-ray14}
\end{table}

 \clearpage
\section{Visualization Results}
\begin{figure}[b]
    \centering
    \includegraphics[width=\textwidth]{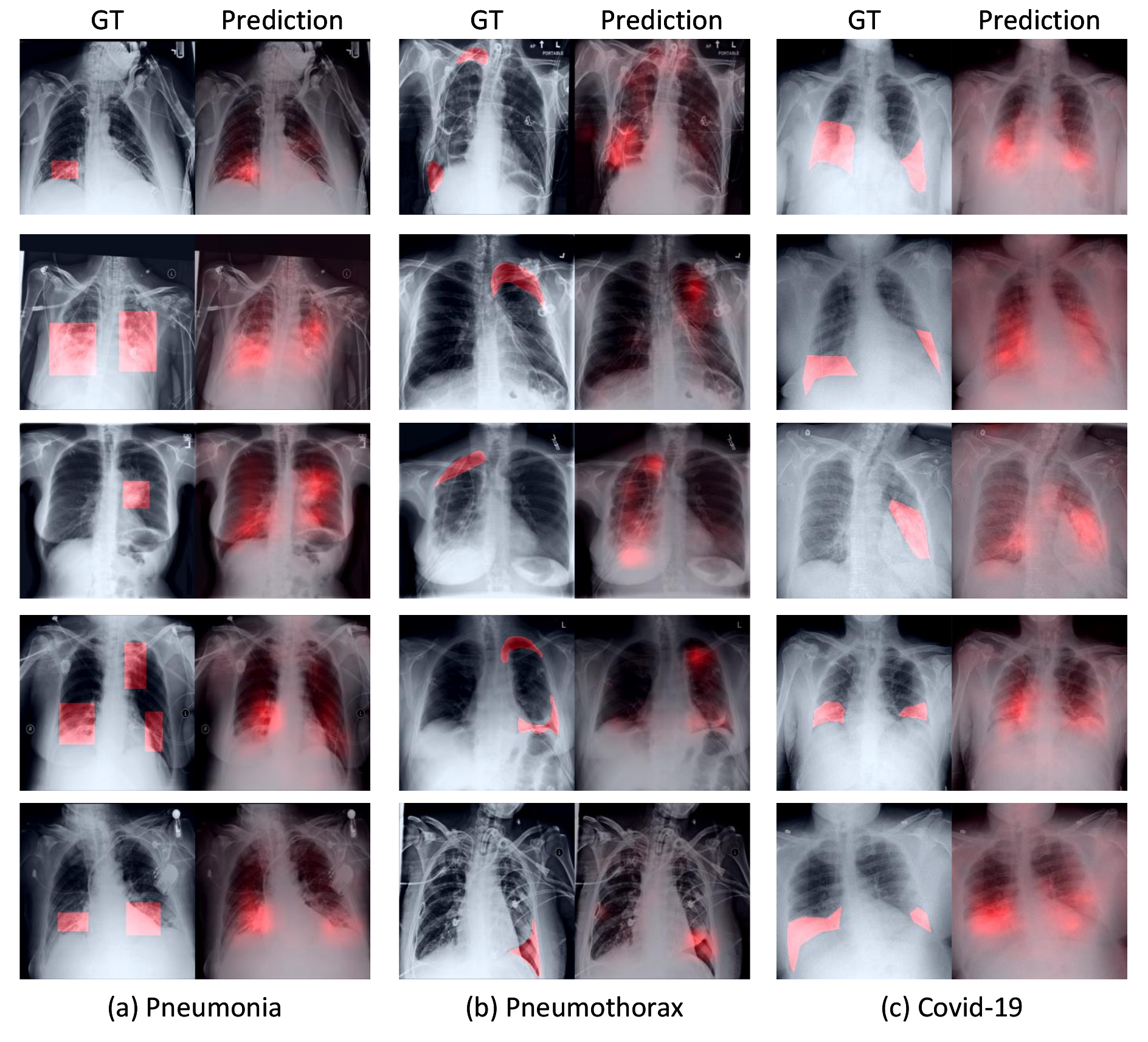}
    \vspace{-0.8cm}
    \caption{The visualization of zero-shot grounding results of our method. Each column represents the results on one disease and the left in it is the ground-truth and right is the heatmap predication of our model. The brighter the red on the figure, the more likely the model considering this region to be associated with abnormalities.}
    \label{fig:visual}
\end{figure}
\label{Visualization}
Fig.~\ref{fig:visual} shows visualization results of our model on zero-shot grounding task. As shown in figure, the ground truth of ``Pneumonia'' is given by bounding box and generally related to a large area region. Thus the metrics on this are higher than other two datasets. Our network captures its regions very well. For ``Pneumothorax'', its abnormality pattern is different from other diseases, which aim to capturing the collapsed part of the lung, rendering darker areas on the images rather than brighter opacity. Its ground-truth masks are generally thin and narrow while our network can still highlight its location. For ``Covid-19'', its image textual was similar to ``Pneumonia'', but since this is a totally new disease, grounding its regions is much more challenging. It requires the model to build relationships between them based on their complex definition and symptoms. The visualization results suggest that our model successfully achieve this, supporting that, for other unseen diseases, our model can also understand their complex descriptions.

\end{document}